\documentstyle[12pt]{article}

\title{
\begin{flushright}
{\normalsize Yaroslavl State University\\
             Preprint YARU-HE-93/01\\
             hep-ph/9404260} \\[3cm]
\end{flushright}
On collinearization of quarks in the quark-gluon decays
of heavy orthoquarkonia}

\author{A.Ya.~Parkhomenko and A.D.~Smirnov \\
{\small\it Division of Theoretical Physics, Department of Physics,}\\
{\small\it Yaroslavl State University, Sovietskaya 14, 150000 Yaroslavl,}\\
{\small\it Russian Federation}}

\date{23 August 1993}

\begin{document}

\maketitle

\begin{abstract}

The decay of heavy orthoquarkonium into quark-antiquark pair and two
gluons is considered. The differential probability of the decay
in the tree approximation is calculated and the probability distributions
of quarks and gluons are obtained. The collinear enhancement
of the bottomonium decay is shown to take place at the u\={u}--,
d\={d} and s\={s}-- pair production and to be absent at the
c\={c}-- production. Some other peculiarities of the considered decays
are discussed.

\end{abstract}

\newpage

\indent The decays of heavy quarkonia give  the  useful  information
on the dynamics of quarks and gluons and  on  the  processes  of  jet
production of hadrons. In  particular  the  many-particle  decays
such as $^{1}S_{0} \rightarrow$  3g, q\={q}g  and $^{3}S_{1}
\rightarrow$ 4g, q\={q}gg are of great interest as
the processes  giving the immediate information on  q\={q}g-- and 3g--
interactions. The nonabelian nature of  ggg-interaction  manifests
itself, for example, in the distribution in invariant  masses  of
two particles$^{1,2}$, as the acomplanarity of four particle decays$^3$,
as the collinearization of the gluons$^4$ and as some other
effects. It is worth noting that the  four particle  decays
of heavy  orthoquarkonia  are  more  available  for  experimental
investigations because of  rather  great  probability  of  direct
production  of orthoquarkonia  in $e^{+}e^{-}$-- and $p\bar{p}$--
collisions. By this reason the detail theoretical analysis of  these
decays is interesting in order to find the optimal conditions
for their experimental observation.

In this work we consider the decay of heavy  orthoquarkonium
with  spin-parity $J^{PC}=1^{- -}$  into  quark-antiquark  pair  and  two
gluons.  The  amplitude  and  differential  probability  of  this
process  in  tree  approximation  and  some  angle   and   energy
distributions of quarks and gluons are obtained and discussed.

The process $^{3}S_{1}(Q\bar{Q}) \rightarrow$  q\={q}gg is described
in the tree approximation by six diagrams shown in fig.1. The amplitude of
this process with the neglect of the relative momentum of the heavy initial
quarks may be presented in the form:

\begin{eqnarray}
M^{\alpha\beta}_{ab} \big ( ^3S_1 (Q\bar{Q}) \rightarrow q{\bar q}gg \big ) =
{ d_{abc} (t_c)_{\alpha\beta} \over 4 \sqrt N_c} \;
{g^4_{st} \psi(0) \over 4 \sqrt{\varepsilon_1 \varepsilon_2 \omega_1 \omega_2}}
\; {J_\mu j_\mu \over p^2 (pk) (Pk_1) (Pk_2)},
\end{eqnarray}
\begin{eqnarray}
J_\mu & = & \bar u_Q(-P/2) \big \lbrace \hat e_2 \big ( - \hat{P}/2 + \hat k_2
+
m \big ) \gamma_\mu \big ( \hat{P}/2 - \hat k_1 + m \big ) \hat e_1 \nonumber
\\
& \times & \big ( ( k_1 + k_2 ) p \big ) + \ldots \big \rbrace u_Q(P/2), \\
j_\mu & = & \bar u_q(p_1) \gamma_\mu u_q(-p_2),
\end{eqnarray}


\noindent where $d_{abc}(a,b,c = 1,2,\ldots,N^2_c-1)$ are symmetrical
constants of $SU(N_{c}), t_{c}$ - generators of $SU(N_{c})$ group normalized
as $Sp(t_{a}t_{b}) = \delta_{ab}/2, \psi(0)$ - nonrelativistic wave
function of quarkonium in the coordinate space, $g_{st}$ -- color charge,
$P, p_i = ( \varepsilon_i, \vec p_i ), k_i = ( \omega_i, \vec k_i ),
i = 1,2$ are 4-momenta of the quarkonium,
final quark and antiquark and gluons correspondingly, $p = p_1 + p_2,
k = k_1 + k_2$, $m$ - mass of the initial quark, $u_Q, u_q$ -- bispinors
of the initial and final quarks. The dots in (2) imply five permutations
of pairs $( \hat e_1; k_1 ), ( \hat e_2; k_2 ), ( \gamma_\mu; p )$.

The corresponding to (1) -- (3) differential probability averaged over
three spin states of initial orthoquarkonium and summed over polarizations
and colors of final particles may be presented in the form:

\begin{eqnarray}
dW = {F_c \over 6 \pi^4} \; {\alpha^4_s \mid \psi(0) \mid^2 Q_{\mu \nu}
G_{\mu \nu} \over \big [ p^2 (kp) (Pk_1) (Pk_2) \big ]^2} \; \;
\delta (P-p-k) \; {d\vec{p_1}d\vec{p_2}d\vec{k_1}d\vec{k_2} \over
\varepsilon_1 \varepsilon_2 \omega_1 \omega_2},
\end{eqnarray}

\noindent where $F_c = ( N^{2}_{c} - 4 ) ( N^{2}_{c} - 1 ) /
( 32 N^{2}_{c} )$ is a color factor of $SU(N_{c})$ -- group, for
$SU(3) F_{c}= 5/36, \alpha_s = g^2_{st} / 4\pi$ is the strong coupling
constant and

\begin{eqnarray}
Q_{\mu \nu} & = & \frac{1}{4} \sum_{pol} j_\mu j^*_\nu = p_{1 \mu} p_{2 \nu} +
p_{1 \nu} p_{2\mu} - { p^2 \over 2} g_{\mu \nu}, \\
G_{\mu \nu} & = & \frac{1}{8 m^2} \sum_{pol} J_\mu J^*_\nu = \big ( p^2
g_{\mu \nu} - p_\mu p_\nu \big ) \bigg \lbrace  \big ( k_1 k_2 \big )^2
\bigg [ {\big( 4 m^2 - k^2 \big)^2 \over 8 m^2} - p^2 \bigg ] \nonumber \\
& + & \big ( p k_1 \big ) \big ( p k_2 \big ) \bigg [ 2 \big ( k_1 k_2 \big )
- {\big ( p k_1 \big ) \big ( p k_2 \big ) \over 2 m^2} \bigg ] \bigg \rbrace
\\
& - & \big ( 4 m^2 - k^2 \big ) l_\mu l_\nu
- \big ( k_1 k_2 \big )^2 \big ( T^{1 1}_{\mu \nu} + T^{2 2}_{\mu \nu} \big )
\nonumber \\
& - & {1 \over 2}
\big ( 4 m^2 - k^2 \big ) \bigg [ \big ( k_1 k_2 \big ) + {\big ( P k_1 \big )
\big ( P k_2 \big ) \over 2 m^2} \bigg ] T^{1 2}_{\mu \nu}, \nonumber
\end{eqnarray}

\noindent with

\begin{eqnarray}
l_\mu & = & \big ( p k_1 \big ) k_{2 \mu} - \big ( p k_2 \big ) k_{1 \mu},
\nonumber \\
T^{i j}_{\mu \nu} & = & 2 \big ( p k_i \big ) \big ( p k_j \big ) g_{\mu \nu}
+ p^2 \big ( k_{i \mu} k_{j \nu} + k_{j \mu} k_{i \nu} \big ) \nonumber \\
& - & \big ( p k_i \big ) \big ( p_\mu k_{j \nu} + p_\nu k_{j \mu} \big ) -
\big ( p k_j \big ) \big ( p_\mu k_{i \nu} + p_\nu k_{i \mu} \big ). \nonumber
\end{eqnarray}

\noindent The expressions for $Q_{\mu \nu}$ and $G_{\mu \nu}$ obtained by
us have the structure such as that of the tensors $L_{\mu \nu}, H_{\mu \nu}$
of Ref.5.

Integrating (4) over quark and antiquark momenta we obtain the probability
distribution in the energies $x_{i} = \omega_{i}/m$ of the gluons and angle
$\vartheta_{g}$ between their momenta in the rest frame of the quarkonium:

\begin{eqnarray}
{d^3W \over dx_1 dx_2 d\cos\vartheta_g} = F_c \;
{\alpha_s^4 \mid \psi(0) \mid^2 \over 36 \pi m^2} \; {F_g \over \eta_g \xi^2_g}
\Big ( 1 + {2 \mu^2 \over \eta_g} \Big ) \sqrt{1- {4 \mu^2 \over \eta_g}} ,
\end{eqnarray}
\begin{eqnarray}
F_g & = & 8 x_1 x_2 \big [ 12 \big ( 1 + \cos^2 \vartheta_g \big )
- 8 \big ( x_1 + x_2 \big ) \big ( 1 - \cos \vartheta_g + \cos^2
\vartheta_g \big ) \nonumber \\
& + & 4 \big ( 1- \cos \vartheta_g \big ) \big [ 2 \big ( x_1 + x_2 \big )^2
- x_1 x_2 \big ( 1 - \cos \vartheta_g - \cos^2 \vartheta_g \big ) \big ]
\nonumber \\
& - & 8 x_1 x_2 \big ( x_1 + x_2 \big ) \big ( 1 - \cos^2 \vartheta_g \big )
+ x^2_1 x^2_2 \big ( 1- \cos \vartheta_g \big )^3 \big ( 3 - \cos \vartheta_g
\big ) \big ], \nonumber \\
\eta_g & = & (P-k)^2 / m^2 = 4 \big ( 1 - x_1 - x_2 \big ) + 2 x_1 x_2
\big ( 1 - \cos \vartheta_g \big ), \nonumber \\
\xi_g & = & \big ( (Pk) - k^2 \big ) / m^2 = 2 \big ( x_1 + x_2 \big )
- 2 x_1 x_2 \big ( 1 - \cos \vartheta_g \big ), \nonumber
\end{eqnarray}

\noindent where $\mu = m_q / m$ is the mass ratio of final and initial quarks,
$P = ( 2m, \vec 0 )$.

Integrating $G_{\mu \nu}$ (6) over the momenta of the gluons we find:

\begin{eqnarray}
F_{\mu \nu} & \equiv & \int \frac{G_{\mu \nu}}{(Pk_1)^2 (Pk_2)^2} \; \delta
\big ( k - k_1 - k_2 \big ) \frac{\vec{dk_1} \vec{dk_2}}{\omega_1 \omega_2}
\nonumber \\
& = & \frac{\pi}{2 z v^4} \bigg \lbrace - f_1 \bigg ( g_{\mu \nu}
- \frac{p_\mu p_\nu}{p^2} \bigg ) + f_2 \; \frac{q_\mu q_\nu}{p^2 m^4}
\bigg \rbrace ,
\end{eqnarray}

\noindent where

\begin{eqnarray}
q_\mu & = & p^2 P_\mu - (Pp) p_\mu , \nonumber \\
f_i & = & g_i + h_i \frac{1 - v^2}{2 v} \ln \frac{1 + v}{1 - v} , \nonumber \\
g_1 & = & v^2 z \big [ - v^8 z^4 + 4 v^6 z^2 ( z^2 + 1 ) - 2 v^4 ( 3 z^4 -
8 z^3 +14 z^2 -24 z + 32 ) \nonumber \\
& + & 4 v^2 ( x^4 - 8 z^3 + 19 z^2 - 24 z +32 ) - z ( z^3 - 16 z^2 + 52 z
- 48 ) \big ] , \nonumber \\
h_1 & = & v^2 z \big [ - v^8 z^4 + 2 v^6 z^2 ( 3 z^2 + 2 ) - 4 v^4 ( 2 z^4 -
4 z^3 - 3 z^2 - 12 z + 16 ) \nonumber \\
& + & 2 v^2 ( z^4 - 2 z^2 - 32 ) + z ( z^3 - 16 z^2 + 52 z - 48 ) \big ] ,
\nonumber \\
g_2 & = & - v^8 z^3 + 20 v^6 z + 2 v^4 ( 3 z^3 - 24 z^2 + 2 z - 8 ) \nonumber
\\
& - & 4 v^2 ( 2 z^3 - 24 z^2 + 45 z - 24 ) + 3 ( z^3 - 16 z^2 + 52 z - 48 ) ,
\nonumber \\
h_2 & = & - v^8 z^3 + 2 v^6 z ( z^2 + 10 ) - 4 v^4 ( z^3 - 4 z^2 + z + 4 )
\nonumber \\
& + & 2 v^2 z ( 3 z^2 - 32 z + 38 ) - 3 ( z^3 - 16 z^2 + 52 z - 48 ) ,
\nonumber \\
z & = & \frac{(Pk)}{2 m^2} , \qquad v = \sqrt{1 - \frac{4 m^2 k^2}{(Pk)^2}} .
\nonumber
\end{eqnarray}

\noindent Here z and v are the energy and the center-of-mass velocity of
the gluon pair in the rest frame of quarkonium.

Using (4), (5) and (8) we obtain the probability distribution in the
energies $y_i = \varepsilon_i / m$ of quarks and angle $\vartheta_q$
between their momenta in the rest frame of the quarkonium:

\begin{eqnarray}
{d^3W \over dy_1 dy_2 d\cos\vartheta_q} = F_c \; {\alpha_s^4 \vert \psi(0)
\vert^2 \over 36 \pi  m^2} \; {F_q \over \eta^2_q \xi^2_q}
\sqrt{\big ( y_1^2 - \mu^2 \big ) \big ( y_2^2 - \mu^2 \big )} ,
\end{eqnarray}

\noindent where

\begin{eqnarray}
F_q & = & \frac{12}{\pi} \, \frac{F_{\mu \nu} Q_{\mu \nu}}{m^2} =
\frac{6}{z v^4} \big [ f_1 ( \eta_q + 2 \mu^2 ) + 2 f_2 \eta_q
( 4 y_1 y_2 - \eta_q) \big ] ,
\nonumber \\
\eta_q & = & p^2 / m^2 = 2 \Big [\mu^2 + y_1 y_2 - \cos \vartheta_g
\sqrt{\big ( y_1^2 - \mu^2 \big ) \big ( y_2^2 - \mu^2 \big )} \Big ],
\nonumber \\
\xi_q & = & \big ( (Pp) - p^2 \big ) / m^2 = 2 \big ( y_1 + y_2 \big ) -
\eta_q, \nonumber \\
z & = & 2 - y_1 - y_2 , \qquad
v = \frac{\sqrt{( y_1 + y_2 )^2 - \eta_q}}{z}, \nonumber
\end{eqnarray}


It's convenient for the further analysis to use the distributions (7), (9)
normalized as

\begin{eqnarray}
f_g ( x_1 , x_2 , \cos\vartheta_g ) = \frac{1}{\alpha_s W_{3g}} \quad
\frac{d^3W}{dx_1 dx_2 d\cos\vartheta_g} ,
\nonumber \\
f_q ( y_1 , y_2 , \cos\vartheta_q ) = \frac{1}{\alpha_s W_{3g}} \quad
\frac{d^3W}{dy_1 dy_2 d\cos\vartheta_q} , \\
W_{3g} = 2 F_c \quad \frac{16 \big ( \pi^2 - 9 \big )}{9} \quad
\frac{\alpha^3_s \mid \psi(0) \mid^2}{m^2} , \nonumber
\end{eqnarray}

\noindent where $W_{3g}$ -- the three-gluonic decay probability of
orthoquarkonium.

The analysis of the quark distribution $f_q$ shows that the probability
of the production of light q\={q}-- pair and two gluons can significantly
increase as the angle between quarks decreases. This collinear effect
has a simple origin  -- the decrease of the denominator in the propagator
of the virtual gluon as the angle between light quarks decreases and
essentialy depends on the mass ratio of the final and initial quarks.
As an example we present the quark distribution $f_q$ as a function of
$\vartheta_q$ and $\mu$ at $y_1 = y_2 = y = 0.4$ in Fig.2. Here one can
see that the quark collinear effect becomes significant at $\mu \le 0.15$
and $\vartheta_q \le 30^\circ$.

Taking the masses of the quarks into account we conclude that the quark
collinear effect takes place in all the quark-gluon decays of the
orthocharmonium. As concerning the quark-gluon decay of the orthobottomonium
the quark collinear effect takes place in the decays with production of
u\= u-, d\= d- and s\= s- pairs only but is absent in the decay with
production of more heavy c\= c- pair because of the rather great mass ratio
$\mu = m_c / m_b \sim 0.3$ (see the peak at small angles in Fig.3 and its
absence in Fig.4). At $y \ge 0.6$ the region of the small angles becomes
kinematicaly forbidden and the collinear effect dissapeares.

By the production of hard ($y \ge 0.8$) q\= q- pair with large angle
between q- and \= q- quarks the energy of two accompanying gluons is small
and the soft gluon enhancement of the q\= q- production occurs in this case
(see the quark curves c in Fig.3,4).

The analysis shows that in contrast with the quark distribution $f_q$
there is no collinear enhancement in the gluon distribution $f_g$, there is
only the soft quark enhansement of the production of the hard gluons
accompanied by the light q\= q- pair (see gluon curve c in Fig.3).

In conclution we resume the main results of the work:
\begin{enumerate}
\item The decays $^3S_1 (Q\bar{Q}) \rightarrow q\bar{q}gg$ of heavy
      orthoquarkonia are considered and the probability distributions
      in the gluon and in the quark variables are found.
\item The enhancement of these decays caused by the collinearization
      of the final quarks is discussed in the mass ratio of the final
      and initial quarks dependence. It is shown that this effect is to take
      place at the angles $\le 30^\circ$ in all the decays $^3S_1 (c\bar{c})
      \rightarrow q\bar{q}gg$ of charmonium, in the decays $^3S_1 (b\bar{b})
      \rightarrow q\bar{q}gg$ of bottomonium with the production of
      u\={u}--, d\={d}-- and s\={s}-- pairs but to be absent in the
      decay $^3S_1 (b\bar{b}) \rightarrow c\bar{c}gg$.
\item The collinear enhancement in the gluon distribution is shown to be
      absent in these decays.
\item The production of hard quarks (gluons) accompanied by two soft gluons
      (quarks) is shown to be enhanced in these decays too.
\end{enumerate}

The quark and gluon distributions obtained by us and their peculiarities
discussed here may be useful for the experimental searches and investigations
of the four-jet events from the decays of heavy orthoquarkonia.

\vskip 1.5em
{\Large\bf Acknowledgment}
\vskip 1em

This work was supported, in part, by a Soros Humanitarian Foundations Grant
awared  by the American Physical Society.

\vskip 1.5em
{\Large\bf References}
\vskip 1em

\begin{enumerate}

\item K.Koller, K.H.Streng, T.F.Walsh and P.M.Zerwas,
      {\it Nucl.Phys.} {\bf B206}, 273 (1982).

\item K.H.Streng,
      {\it Z.Phys.} {\bf C27}, 107 (1985).

\item T.Muta and T.Niuya,
      {\it Progr.Theor.Phys.} {\bf 68}, 1735 (1982).

\item A.D.Smirnov,
      {\it Yad.Fiz.} {\bf 47}, 1380 (1988).

\item L.Clavelli, P.H.Cox and B.Harms,
      {\it Phys.Rev.} {\bf D31}, 78 (1985).

\end{enumerate}

{\Large\bf Figure captions}

\begin{enumerate}

\item Diagrams of the decay $^3S_1 (Q\bar Q) \rightarrow q\bar{q}gg$
      in the tree approximation.

\item The quark distribution $f_q$ as a function of $\vartheta_q$ and
      mass ratio $\mu$ at $y_1 = y_2 = y = 0.4$.

\item Angle distributions of gluons at $x_1 = x_2 = x$ and quarks at
      $y_1 = y_2 = y$ in the decay $^3S_1 (b\bar b) \rightarrow
      s\bar{s}gg$: a) $x = 0.4$ or $y = 0.4$; b) $x = 0.6$ or $y = 0.6$;
      c) $x = 0.8$ or $y = 0.8$.

\item Angle distributions of gluons at $x_1 = x_2 = x$ and quarks at
      $y_1 = y_2 = y$ in the decay $^3S_1 (b\bar b) \rightarrow
      c\bar{c}gg$: a) $x = 0.4$ or $y = 0.4$; b) $x = 0.6$ or $y = 0.6$;
      c) $y = 0.8$.

\end{enumerate}

\end{document}